\documentstyle[editedvolume,numreferences,epsfig,epsf]{crckapb} 

\begin{opening}
\title{NOISE CORRELATIONS, ENTANGLEMENT,\protect\\
       AND BELL INEQUALITIES}

\author{T. MARTIN AND A. CREPIEUX}
\institute{CPT et Universit\'e de la M\'editerran\'ee\\
           Case 907, 13288 Marseille}
\author{N. CHTCHELKATCHEV}
\institute{L.D. Landau Institute for Theoretical Physics\\
           Kosygina Str. 2, 117940 Moscow}
\end{opening}

\runningtitle{NOISE CORRELATIONS AND BELL}

\begin{document}


\section{Introduction}

In condensed matter physics, the interactions between 
the constituents of the system are typically known. The  
building blocks 
are mere electrons, protons and neutrons, their
collective  behavior has been shown to lead to a variety of 
astonishing phenomena. Classic examples of such quantum
correlated systems are superconductivity \cite{de_gennes},
superfluidity \cite{kalatnikov} and the 
fractional quantum  Hall effect (FQHE) \cite{laughlin}.
In these instances, the departure from  usual
behavior is often symptomatic of  the presence of a non
trivial ground state: a ground  state which cannot be described
by a systematic  application of perturbation theory
on the noninteracting system. 

Investigations on such ground states naturally lead to
that of the elementary excitations of the system.
One typically probes the system with an external
interaction which triggers the population of excited 
states. The subsequent measurement of the 
thermodynamical properties then provides some crucial
information. From a different angle,
transport measurements deal with open systems connected 
to reservoirs. There are several ways to approach the issue of
(quasi)-particle correlations using transport experiments.
One can chose to study a system where interactions are most
explicit, such as the FQHE, otherwise, to look for statistical
interactions in Fermi and Bose systems.
Noise being a multi-particle diagnosis, it can
be employed to study both issues.

The aim of this chapter is to describe two situations where
positive noise correlations
can be directly monitored using a transport
experiment, either with a superconductor or with a correlated 
electron system.  To be more precise, 
the present text reflects the presentations
made by the three authors during the Delft NATO workshop. 
Bell inequalities and quantum mechanical non-locality 
with electrons injected from a superconductor will be
addressed first \cite{chtchelkatchev}.
Next, noise correlations will be computed in a carbon 
nanotube where electrons are injected in the bulk 
from a STM tip \cite{crepieux}. 
The first topic 
is the result of an ongoing collaboration with G. 
Lesovik and G. Blatter over the years. 
The unifying theme is that in both branched quantum circuits,  
entanglement is explicit and can be illustrated via noise 
correlations.
Entanglement can be achieved either for pairs of electrons
in the case of superconductor sources connected to Fermi 
liquid leads, or alternatively for pairs of quasiparticle
excitations of the correlated electron fluid.  

A normal metal fork attached to a superconductor 
can exhibit positive correlations 
\cite{torres_martin,lesovik_martin_blatter},
which had been attributed primarily to photonic systems in
the seminal Hanbury-Brown and Twiss experiment \cite{hbt}. They
arise when the source of particles is a superconductor 
\cite{torres_martin,gramespacher,belzig,schechter,taddei,bouchiat}
or they also occur in systems with floating
voltage probes \cite{texier}. Here, evanescent Cooper pairs can
be emitted on the normal side, due to the proximity effect
\cite{proximity}. These Cooper pairs can either decay in one
given lead, which gives a negative contribution to noise
correlations, or may split at the junction on the normal side
with its two constituent electrons propagating in different
leads. This latter effect constitutes a justification for
positive noise correlations. 
If filters, such as quantum dots, are added to the leads, 
such a mechanism generates delocalized, entangled electron 
pairs \cite{lesovik_martin_blatter,recher_sukhorukov_loss}.
In order to exit from the superconductor, a Cooper pair 
has to be split between the two normal leads. 
This provides a solid state analog of 
Einstein-Podolsky-Rosen (EPR) states which were proposed to    
demonstrate the non-local nature of quantum mechanics \cite{EPR}.
Photon entanglement has triggered the proposition of new
information processing schemes based on quantum mechanics for
quantum cryptography or for teleportation \cite{bouw_book}. 
Concrete proposals for quantum
information prossessing devices 
based on electron transport and electron interactions
in condensed matter have been recently presented.
\cite{lesovik_martin_blatter,recher_sukhorukov_loss,recher_loss,divicenzo}.
Here we will analyze whether a non-locality test can be
conceived for electrons propagating in quantum wave guides.
It is indeed possible to perform the exact analog of a Bell
inequalities violation \cite{bell} for photons in a condensed
matter system \cite{chtchelkatchev}.

As cited above, positive correlation do not necessarily 
require a superconducting source of electrons. 
A particular geometry using a one dimensional
correlated electron liquid can be studied for the 
same purposes. In particular, it
allows to probe directly the underlying charges of the
collective excitations in the Luttinger liquid. 
In order to make contact with experiments, the setup
consists of a nanotube whose bulk is contacted by an STM tip
which injects electrons by tunneling, 
while both extremities of the nanotube
collect the current.  The current, the noise and the noise
correlations are computed, and the effective charges are
determined by comparison with the Schottky formula. For an
``infinite'' nanotube, the striking result is that noise
correlations contribute to second order in the electron
tunneling, in sharp contrast with a fermionic 
system which requires fourth order. The noise correlations are
then positive, because the tunneling electron wave function is
split in two counter propagating modes of the collective
excitations in the  nanotube.

The transport properties of
quasiparticles will be addressed from first principles. The
understanding of interactions -- statistical or otherwise -- in
noise correlations experiments will be approached here from the
point of view of scattering theory and of the Keldysh technique
\cite{datta_imry}. 

\section{Hanbury--Brown and Twiss correlations}

Particularly interesting is the role of electronic correlations
in quantum transport. Correlations can have several causes.
First, they may originate from the interactions between the
particles themselves. Second, correlations are generated by a 
measurement which involves two or more particles. In the latter
case, non-classical correlations may occur solely because of the 
bosonic or fermionic statistics of particles, with or without 
interactions.
The measurement of noise -- the Fourier transform of the
current--current correlation function -- constitutes a two
particle measurement, as implied in the average of the two 
current operators:
\begin{equation}
S_{\alpha\beta}(\omega=0)=
{1\over 2}
\int dt 
\left(\langle I_\alpha(t)I_\beta(0) +
I_\beta(0)I_\alpha(t)\rangle - 2\langle I_\alpha\rangle
\langle I_\beta\rangle\right)~.
\label{noisenoise}\end{equation} 
Here $I_\alpha$ is the current 
operator in reservoir $\alpha$, and the time arguments on the
average currents have been dropped, assuming a stationary 
regime. 

Consider the case of photons propagating in vacuum: the
archetype of a weakly interacting boson system. It was  shown
\cite{hbt} that when a photon beam is extracted from a thermal
source such as a mercury arc lamp, the intensity correlations
measured in two  separated photo-multipliers are always positive. 
On average, each photon scattering state emanating from the
source  can be populated by several photons at a time -- due to
the bunching property of bosons. As a result when a photon is
detected in one of the photo-multipliers, it is likely 
to be correlated with another detection in the other
photo-tube. The positive correlations can be considered as a 
diagnosis of the statistics of the carriers performed with a
quantum transport  experiment. 

\subsection{Noise correlations in normal metals} 

\begin{figure}
\epsfxsize 8.5 cm
\centerline{\epsffile{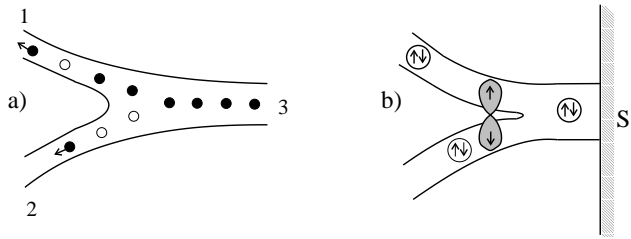}}
\medskip
\caption{\label{fig1}
a) Hanbury-Brown and Twiss geometry in a normal metal
fork with electrons injected from $3$ and collected in
reservoirs $1$ and $2$. Occupied (empty) electron wave packet
states  are identified as black (white) dots.
b) Hanbury-Brown and Twiss geometry in a superconductor--normal
metal fork. Cooper pairs are emitted from the superconductor,
and the two constituent electrons can either propagate in the
same lead, or propagate in an entangled state in both 
leads.} 
\end{figure}

What should be the equivalent test for electrons ? 
A beam of electrons can be viewed as a train of
wave packets, each of which is populated at most by two
electrons with opposite spins. If the beam is fully occupied, 
negative correlations are expected because the measurement of an
electron in one detector is accompanied by the absence of a
detection in the other one, as depicted in Fig. \ref{fig1}a.
It was understood recently \cite{martin_landauer} that if
the electrons propagate in a quantum wire with few lateral modes,
maximal occupancy could be reached, and the anti-correlation
signal  would then be substantial. 

Consider the
device drawn  in Fig. \ref{fig1}a : electrons emanating from
reservoir $3$ have a probability amplitude $s_{1(2)3}$ to end up
in reservoir $1(2)$. 
For simplicity we assume that each lead is connected to a
single electron channel.
The scattering matrix describing this
multi-terminal system is hermitian because of current
conservation. 
On general grounds, it is possible to derive the following sum
rule for the autocorrelation noise and the noise correlations
\cite{buttiker_prb}:
\begin{equation}
\sum_\alpha S_{\alpha\alpha} (\omega =0)+
\sum_{\alpha\neq\beta} S_{\alpha\beta} (\omega
=0)=0~.\end{equation} 
Note that this sum rule only holds for normal
conductors, but it not valid for conductors which involve
superconducting reservoirs. The
scattering theory of electron transport \cite{datta_imry} then
specifies how to compute the current and the noise
correlations between the two  branches. The current in lead
$\alpha$ is given by  \begin{equation}
\langle I_\alpha \rangle = {e\over h}\int dE~
\left({\rm
Tr}[{\bf 1}_\alpha -
s^\dagger_{\alpha\alpha}s_{\alpha\alpha}]f_\alpha(E)-
\sum_{\beta\neq\alpha}  
s^\dagger_{\alpha\beta}s_{\alpha\beta}f_\beta(E)
\right)
~,\end{equation}~
with $s_{\alpha\beta}$ the amplitude to go from reservoir
$\beta$ to reservoir $\alpha$, ${\bf 1}_\alpha$ is the identity 
matrix for lead $\alpha$, and $f_\alpha$ the associated
Fermi function.  The zero frequency noise
correlations in the zero temperature limit are found in general
to be \cite{buttiker_prb}: 
\begin{equation}
S_{\alpha\beta}
= 
{e^2\over h} \sum_{\gamma\neq\delta}\int dE~{\rm Tr}
[s^\dagger_{\alpha\gamma}s_{\alpha\delta} 
s^\dagger_{\beta\delta}s_{\beta\gamma}]f_\gamma(E)
[1-f_\delta(E)]~.
\end{equation}
Note that when considering the tunnel limit, 
the lowest non vanishing contribution is of fourth order in the
tunneling amplitude $\Gamma\sim  s_{\alpha\beta}$ for fermionic
particles. Applying the above results to the three-terminal
situation depicted in Fig. \ref{fig1}  
in the presence of a symmetric voltage bias $V$
between $3$ and $1(2)$ so that
no flow occurs between $1$ and $2$:
\begin{equation} 
S_{12}(0)= -{2e^3|V|\over h} |s_{13}s_{23}|^2~.
\label{electron_correlations} 
\end{equation}  
These electronic
noise correlations were measured recently \cite{henny}
by two groups working either in the integral quantum Hall effect regime
or in the ballistic regime, 
with beam splitters designed with metallic gates. 

Negative correlations here are most natural,
because the injection of electrons is made from a degenerate
Fermi gas. 
Yet there exist situations where they can be
positive in a fermionic system. 

\subsection{Fork geometry with a superconductor source}

If the reservoir which injects electrons
in the fork is a superconductor as in Fig. \ref{fig1}b, both
positive and negative noise correlations are possible
\cite{torres_martin}. 
Charge transfer between the injector and
the two collectors $1$ and $2$ is then specified by the Andreev
scattering  process, where an electron is reflected as a hole.
Positive correlations are linked to
the proximity effect, as superconducting correlations (Cooper
pairs) leak in the two normal leads. 
Depending on the nature of the junction in  Fig.
\ref{fig1}b, it may be more favorable for a pair to  be
distributed among the two arms than for a pair to enter a lead
as a whole. The detection of an electron
in $1$ is then accompanied by the  detection of an electron in
$2$, giving a positive correlation signal. 

In the scattering theory
for normal--superconductor (NS) systems
\cite{ns_transport},
the fermion operators which enter the current operator 
are given in terms of the quasiparticle states 
using the Bogolubov transformation:
$\psi_{\sigma}(x) =\sum_n \left( u_n(x) \gamma_{n \, \sigma} 
- \sigma v_n^*(x) \gamma^{\dagger}_{n \, -\sigma} \right)$.
Here, $\gamma^{\dagger}_{n \, \sigma}$ 
are quasiparticle creation operators,  
$n=(i,\alpha,E)$ contains information on the reservoir ($i$)
from which the particle ($\alpha=e,h$) is incident with energy $E$ 
and $\sigma$ labels the spin.

The contraction of these two operators gives
the distribution function of the particles injected from 
each reservoir, which for a potential bias
$V$ are: $f_{ie}\equiv f(E-eV)$ for electrons incoming 
from $i$, similarly $f_{ih}\equiv f(E+eV)$ for holes,
and $f_{i\alpha}=f(E)$ for both types of quasiparticles  
injected from the superconductor ($f$ is the Fermi--Dirac distribution).
The solution of the Bogolubov--de Gennes equations
provide the electron and hole wave functions describing
scattering states $\alpha$ (particle) and $i$ (lead) are
expressed in terms of the  elements $s_{ij\alpha\beta}$ of the 
S--matrix which describes the whole NS ensemble:

\begin{eqnarray}
u_{i\alpha}(x_j)&=&[\delta_{ij}\delta_{\alpha e}e^{ik_+x_j}+
s_{jie\alpha}e^{-ik_+x_j}]/\sqrt{v_+}~,\\ 
v_{i\alpha}(x_j)&=&[\delta_{ij}\delta_{\alpha h}e^{-ik_-x_j}+
s_{jih\alpha}e^{ik_-x_j}]/\sqrt{v_-}~,
\label{u et v} 
\end{eqnarray}
where $x_j$ denotes the position in normal lead $j$ and $k_\pm$ ($v_\pm$)
are the usual momenta (velocities) of the two branches.

\begin{figure}
\epsfxsize 8.5 cm
\centerline{\epsffile{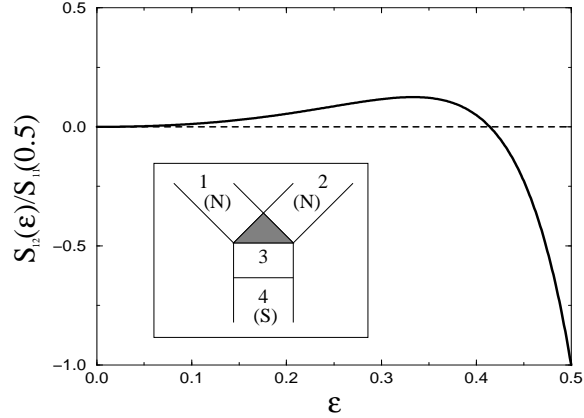}}
\medskip
\caption{\label{fig2}
Noise correlation between the two normal reservoirs of the device
(inset), as a function of the transmission probability of the 
beam splitter, showing both positive and negative correlations.
Inset: the device consists of a superconductor ($4$)--normal ($3$)
interface which is connected by a beam splitter 
(shaded triangle) to reservoirs ($1$) and ($2$). $\epsilon=0.5$
corresponds to maximal transmission.}
\end{figure}

Specializing now to the NS junction connected to a beam splitter
(inset of Fig. \ref{fig2}), $6\times 6$ matrix elements are
sufficient to  describe all scattering processes. At zero
temperature,  the noise correlations between 
the two normal reservoirs simplify to:
\begin{eqnarray}
\nonumber
&&S_{12}(0)= \frac{2e^2}{h}\int_0^{eV} dE \sum_{i=1,2}\\
\nonumber
&&\times\Bigl[
  \sum_{j=1,2}
  \left( s^*_{1iee} s_{1jeh} - s^*_{1ihe} s_{1jhh} \right)
  \left( s^*_{2jeh} s_{2iee} - s^*_{2jhh} s_{2ihe} \right)
\\
&& 
~~~+\sum_{\alpha=e,h} \left( s^*_{1iee} s_{14e\alpha} - s^*_{1ihe} s_{14h\alpha} \right)
  \left( s^*_{24e\alpha} s_{2iee} - s^*_{24h\alpha} s_{2ihe} \right)\Bigr]
~,
\label{correlations splitter}
\end{eqnarray}
where the subscript $4$ denotes the superconducting lead.
The first term represents  
normal and Andreev reflection processes, while the 
second term invokes the transmission of quasiparticles
through the NS boundary. It was noted 
\cite{martin_ns} that in the pure Andreev regime 
the noise correlations vanish when the junction 
contains no disorder: electron (holes) incoming 
from $1$ and $2$ are simply converted into holes
(electrons) after bouncing off the NS interface. 
The central issue, whether changes in the transparency 
can induce changes in the sign of the correlations, is now 
addressed. 

We now consider only the subgap or Andreev regime, 
were $eV\ll \Delta$, the  superconducting gap, for which
a simple model for a disordered NS junction is readily available. 
The junction is composed of four
distinct regions (see inset Fig. \ref{fig2}). The interface between $3$
(normal) and $4$ (superconductor) exhibits only Andreev
reflection, with scattering amplitude  for electrons into holes
$r_A=\gamma\exp(-i\phi)$  (the phase of
$\gamma=\exp[-i\arccos(E/\Delta)]\simeq-i$  is the Andreev phase and
$\phi$ is  the phase of the superconductor). 
Next, $3$ is connected to two reservoirs $1$ and $2$ by a beam 
splitter which is parameterized by a single parameter 
$0<\epsilon<1/2$ identical to that of Ref. \cite{gefen}:
the splitter is symmetric, its scattering matrix coefficients
are real, and transmission between $3$ and the reservoirs
is maximal when $\epsilon=1/2$, and vanishes at $\epsilon=0$. 

Performing the energy integrals in Eq. 
(\ref{correlations splitter}):
\begin{equation}
S_{12}(\epsilon) = \frac{2e^2}{h} eV \frac{\varepsilon^2}{2(1-\varepsilon)^4}
\left( -\varepsilon^2 -2\varepsilon+1 \right)~. 
\end{equation}
The noise correlations vanish at $\epsilon=0$ and 
$\epsilon=\sqrt{2}-1$. 
The correlations (Fig. \ref{fig2}) are positive (bosonic) for 
$0<\epsilon<\sqrt{2}-1$ and negative (fermionic) for 
$\sqrt{2}-1<\epsilon<1/2$. At maximal transmission into the 
normal reservoirs ($\epsilon=1/2$), the correlations normalized
to the noise in $1$ (or $2$) give the negative minimal 
value: electrons and holes do not interfere and propagate
independently into the normal reservoirs. 
It is then expected to obtain the signature of a purely 
fermionic system. 
When the transmission $\epsilon$ is decreased, 
multiple Andreev processes start taking some importance.
Further reducing the 
beam splitter transmission allows to balance the contribution 
of split Cooper pairs with that of Cooper pairs entering the
leads as a whole. 
Note that inclusion of disorder or/and additional leads has been
discussed by many authors, using either the scattering approach
\cite{gramespacher,taddei,bouchiat} or circuit theory
\cite{belzig} to treat the diffusive limit.
To summarize, both positive and negative noise correlations 
are possible there, and such results are discussed
in detail in the contributions of \cite{buttiker_kluwer} and 
\cite{belzig_kluwer} contained in this volume.
In particular, the possibility for positive noise cross-correlation
is reduced for asymmetric multichannel conductors. 
Although sample
specific, the scattering theory results of Refs. \cite{torres_martin,bouchiat}
are found to be rather robust in the presence of disorder. 

\subsection{Filtering spin/energy}

\begin{figure}
  \centerline{\epsfxsize=7.2cm
\epsfysize=6.5cm\epsfbox{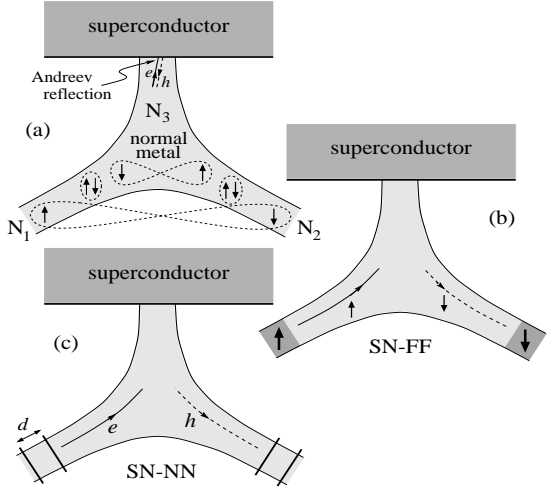}}  
\caption{Normal-metal--superconductor
(NS) junction with normal-metal leads arranged in a fork
geometry.     (a) Without filters, entangled pairs of
quasi-particles (Cooper pairs) injected in N$_3$ propagate into
leads N$_1$ or     N$_2$ either as a whole or one by one.  The
ferromagnetic filters     in setup (b) separates the entangled
spins, while the energy filters in 
(c) separate electron- and hole quasi-particles.\label{fig3}}
\end{figure}

Applying spin or
energy filters  to the normal arms $1$ and $2$ (Fig. \ref{fig3}),
it is possible to generate positive correlations only
\cite{lesovik_martin_blatter}. 
For electron emanating from a superconductors, it is possible
to project either the spin or the energy with an appropriate
filter, without perturbing the entanglement of the remaining
degree of freedom (energy or spin).
Energy
filters, which are more appropriate towards a comparison with
photon experiments, will have resonant energies symmetric above
and below the superconductor chemical potential which serve to
select electrons (holes) in leads $1(2)$. The positive
correlation  signal then reads:   
\begin{equation} 
S_{12}(0)= {e^2\over h} \sum_{\bf \zeta}
\int_0^{e|V|} d\varepsilon
{\cal T}^A_{\bf \zeta}(\varepsilon)
[1-{\cal T}^A_{\bf \zeta}(\varepsilon)] ~,
\label{noise _filters}
\end{equation}
where the index ${\bf \zeta}=h,\sigma, 2$, 
($\sigma=\uparrow,\downarrow$) identifies the incoming hole state
for energy filters (positive energy electrons with arbitrary
spin are injected in lead 1 here).
${\bf \zeta}=h,\uparrow,1 ~(h,\downarrow,2 )$ applies
for spin filters (spin up electrons -- with positive energy -- 
emerging from the superconductor
are selected in lead 1).
${\cal T}^A_{\bf \zeta}$ is then the corresponding (reverse)
crossed-Andreev reflection
probability \cite{melin_feinberg} for each type of setup: 
the energy (spin) degree of freedom 
is frozen, while the spin (energy) degree of freedom 
is unspecified.
$eV<0$ insures that the constituent
electrons of a Cooper pair from the superconductor are emitted
into the leads without suffering from the Pauli exclusion principle.
Moreover, because of such filters, 
the propagation of a Cooper pair as a whole in a
given lead is prohibited. 
Note the similarity with the quantum noise
suppression  mentioned above. This is no
accident: by adding constraints to our system, it has become a
two terminal device, such that the noise correlations between
the two arms  are identical to the noise in one arm:
\begin{equation}
S_{11}(\omega=0)=S_{12}(\omega=0)~.
\end{equation}
The positive correlation and the perfect locking between 
the auto and cross correlations provide a serious symptom of
entanglement. One can speculate (however without 
a rigorous proof yet) that the wave function which
describes the two electron state in the case 
of spin filters reads:
\begin{eqnarray}
|\Phi^{\rm spin}_{\varepsilon,\sigma}\rangle =
\alpha|\varepsilon,\sigma;
-\varepsilon,-\sigma \rangle
+\beta |- \varepsilon,\sigma; 
\varepsilon, - \sigma \rangle~,
\label{wave_function_spin}
\end{eqnarray}
where the first (second) argument in $|\phi_1;\phi_2\rangle$
refers to the quasi-particle state in lead 1 (2) evaluated behind
the filters, $\varepsilon$ is the energy and $\sigma$ is a spin index. The
coefficients  $\alpha$ and
$\beta$ can be tuned by external parameters, 
e.g., a magnetic field.  Note that by projecting the spin
degrees of freedom  in each lead, the spin entanglement is
destroyed, but energy degrees of freedom are 
still entangled, and can help provide 
a measurement of quantum mechanical non locality
nevertheless. A measurement
of energy $\varepsilon$ in lead $1$ (with a quantum dot)
projects the wave function 
so that the energy $-\varepsilon$ has to occur in lead $2$. 
On the other hand, energy filters do  
preserve spin entanglement, and are appropriate to 
make a Bell test (see below). In this case the two 
electron wavefunction takes the form:
\begin{eqnarray}
|\Phi^{\rm energy}_{\varepsilon,\sigma}\rangle =
\alpha|\varepsilon,\sigma;
-\varepsilon,-\sigma \rangle
+\beta |\varepsilon, -
\sigma; - \varepsilon,\sigma \rangle~.
\label{wave_function_energy}
\end{eqnarray}
Electrons emanating from the
energy filters (coherent quantum dots) could be analyzed
provided a measurement can be performed on the 
spin of the outgoing electrons with ferromagnetic leads.


\subsection{Tunneling approach to entanglement}
\label{entanglement_section}

\begin{figure} 
\centerline{\epsfxsize=6.0cm \epsfbox{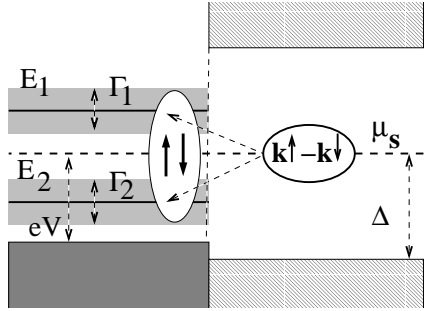}}  
\caption{\label{fig4} Transfer of a Cooper pair on two quantum
energy levels $E_{1,2}$ with a finite width 
$\Gamma_{1,2}$.  The superconductor is located on the right
hand side. The transfer
of a Cooper pair gives an entangled state in the dots
because it implies the creation and destruction of the same
quasiparticle in the superconductor. The source drain voltage
$eV$ for measuring noise  correlations is indicated.}  
\end{figure}

We recall a perturbative argument which supports the claim 
that two electrons originating from the same Cooper pair are
entangled. Consider a system composed of two quantum dots
(energies $E_{1,2}$) next to a superconductor. 
An energy diagram is depicted in Fig. \ref{fig4}.
The electron states  in the superconductor are specified by the
BCS wave function $|\Psi_{BCS}\rangle=\prod_k
(u_k+v_kc^\dagger_{k\uparrow}  
c^\dagger_{-k\downarrow})|0\rangle$. Tunneling to the dots is
described by a single electron hopping Hamiltonian:
\begin{equation} H_T=\sum_{k\sigma}[t_{1k}c^\dagger_{1\sigma}+
 t_{2k}c^\dagger_{2\sigma}] c_{k\sigma}+ h.c.~,
\end{equation} 
 with $c^\dagger_{k\sigma}$ creates an electron with spin $\sigma$.
Now let assume that the transfer Hamiltonian acts on a single
Cooper pair.
 
Using the T-matrix to lowest ($2$nd) order, the
wave function  contribution of the two particle state with one
electron  in each dot reads:
\begin{eqnarray}
|\delta \Psi_{12}\rangle
&=&
H_T{1\over i\eta-H_0}
H_T|\Psi_{BCS}\rangle\nonumber\\
&=&
\sum_k 
v_{k}u_{k}t_{1k}t_{2k}
\left({1\over i\eta
-E_{k}-E_1}+
{1 \over i\eta-E_{k}-E_2}
\right)\nonumber\\
&~&~~\times 
[c^\dagger_{1\uparrow}c^\dagger_{2\downarrow}-
c^\dagger_{1\downarrow}c^\dagger_{2\uparrow}] 
|\Psi_{BCS}\rangle~,
\label{entangled_dots}
\end{eqnarray}
where
$E_k$ is the
energy of a Bogolubov quasiparticle.  The state of Eq.
(\ref{entangled_dots}) has entangled spin degrees of freedom.
This is clearly a result of the spin symmetry of the
tunneling Hamiltonian. Given the nature of the correlated
electron state in the superconductor in terms of Cooper pairs,
$H_T$ can only produce singlet states in the dots.  

\section{Bell inequalities with electrons}
 
In photon experiments, entanglement is identified by a violation 
of Bell inequalities (BI) -- which are obtained with a hidden 
variable theory. But in the case of photons, the BIs have been 
tested using photo-detectors measuring coincidence rates 
\cite{aspect}. 
Counting quasi-particles one-by-one in coincidence measurements
is difficult to achieve in solid-state systems
where stationary
currents and noise 
are the natural observables
\cite{ns_transport}. Here, the BIs are re-formulated in terms 
of current-current cross-correlators (noise correlations) 
\cite{chtchelkatchev}. 

\begin{figure} 
\epsfxsize 8.5 cm
\centerline{\epsffile{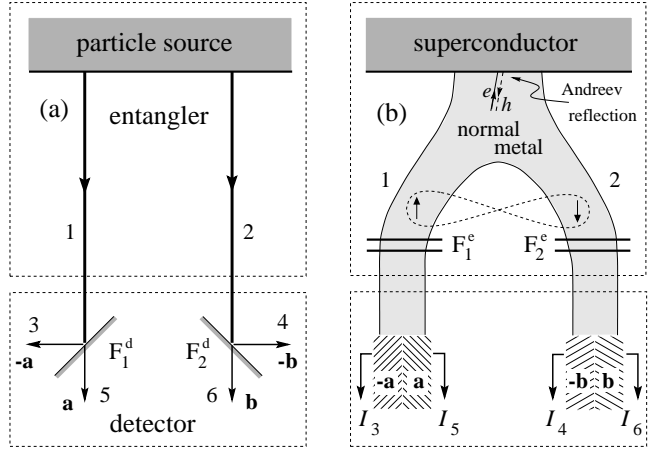}}
\caption{\label{fig5} a) Schematic setup for the measurement of 
Bell inequalities: a source emits particles into leads 1 and 2. 
The detector measures the correlation between beams labelled with 
odd and even numbers. Filters F$^{\rm d}_{1(2)}$ select the spin: 
particles with polarization along the direction $\pm{\bf 
a}(\pm{\bf b})$ are transmitted through filter F$^{\rm d}_{1(2)}$ 
into lead 5 and 3 (6 and 4). b) Solid state implementation, with 
superconducting source emitting Cooper pairs into the leads. 
Filters F$^{\rm e}_{1,2}$ (e.g., Fabry-Perot double barrier 
structures or quantum dots) prevent Cooper pairs from entering a 
single lead. Ferromagnets with orientations $\pm{\bf a},~\pm{\bf 
b}$ play the role of the filters F$^{\rm d}_{1(2)}$ in a); they 
are transparent for electrons with spin aligned along their 
magnetization axis.} 
\end{figure} 

In order to derive Bell inequalities, we consider 
that a source
provides two streams of particles (labeled $1$ and $2$) as in
Fig. \ref{fig5}a injecting quasi-particles  into two arms
labelled by indices 1 and 2. Filter F$^d_{1(2)}$ are transparent
for  electrons spin-polarized along the direction 
${\bf a}$(${\bf b}$). 
Assuming separability and locality \cite{bell} the density 
matrix for joint events in the leads $\alpha,\beta$ is
chosen to be:
\begin{equation}
\rho=\int d\lambda f(\lambda) 
\rho_\alpha(\lambda)\otimes\rho_\beta(\lambda)~,
\label{density_matrix}
\end{equation} 
where the lead 
index $\alpha$ is even and $\beta$ is odd (or vice-versa); the 
distribution function $f(\lambda)$ is positive.
$\rho_\alpha(\lambda)$ are standard density 
matrices for a given lead, which are hermitian. 
The total density matrix 
$\rho$ is the most general density matrix one can build
for the source/detector system assuming no entanglement
and only local correlations \cite{werner}.

Consider the current operator
$I_\alpha(t)$ in lead $\alpha=1,\ldots,6$ (see Fig. \ref{fig5}) and the 
associated particle number operator $N_\alpha(t,\tau) 
=\int_t^{t+\tau} I_\alpha (t^\prime)dt^\prime$.
Particle-number correlators are defined as:
\begin{equation} 
\langle N_\alpha(t,\tau) 
N_\beta(t,\tau)\rangle_\rho=\int d\lambda f(\lambda) \langle 
N_\alpha(t,\tau)\rangle_\lambda\langle 
N_\beta(t,\tau)\rangle_\lambda~, 
\end{equation}
with indices $\alpha/\beta$ 
odd/even or even/odd. The average $\langle 
N_\alpha(t,\tau)\rangle_\lambda$ depends on the state of the 
system in the interval $[t,t+\tau]$. An average over large 
time periods is introduced in addition to 
averaging over $\lambda$, e.g., 
\begin{equation}
   \label{time averaging} \langle
   N_\alpha(\tau)N_\beta(\tau)\rangle
   \equiv \frac 1{2T}\int_{-T}^T dt\langle
   N_\alpha(t,\tau)N_\beta(t,\tau)\rangle_\rho~,
\end{equation}
where $T/\tau\to \infty$ (a similar definition applies to 
$\langle N_\alpha(\tau)\rangle$).  
Particle number fluctuations are written as
$\delta N_\alpha(t,\tau)\equiv 
N_\alpha(t,\tau)-\langle N_\alpha(\tau)\rangle$. 

Let $x,x^\prime,y,y^\prime,X,Y$ be real numbers such 
that:
\begin{equation}
|x/X|,|x^\prime/X|,|y/Y|,|y^\prime/Y|<1~.
\end{equation}
Then
$-2XY\leq xy-xy^\prime+x^\prime
   y+x^\prime y^\prime\leq 2XY$.
Define accordingly:
\begin{eqnarray}
&&   x= \langle N_5(t,\tau)\rangle_\lambda 
   -\langle N_{3}(t,\tau)\rangle_\lambda~,~
   x{^\prime} =
   \langle N_{5^{\prime}}(t,\tau)\rangle_\lambda -\langle
   N_{3^{\prime}}(t,\tau)\rangle_\lambda~,\label{xx'}\\
&&   y=\langle
   N_6(t,\tau)\rangle_\lambda-\langle
   N_4(t,\tau)\rangle_\lambda~, ~
   y^{\prime} =\langle
   N_{6^{\prime}}(t,\tau)\rangle_\lambda-\langle
   N_{4^{\prime}}(t,\tau)\rangle_\lambda~,\label{yy'}
   \end{eqnarray}
where the subscripts with a `prime' indicate a different 
direction of spin-selection in the detector's filter (e.g., let 
${\bf a}$ denote the direction of the electron spins in lead 5 
($-\bf a$ in lead 3), then the subscript $5^{\prime}$ in 
Eqs. (\ref{xx'}) and (\ref{yy'}) 
means that the electron spins in lead 5 are 
polarized along ${\bf a}^{\prime}$ (along $-\bf a^\prime$ in the 
lead 3). The quantities $X,Y$ are defined as 
   \begin{eqnarray}
   X&=& \langle  N_5(t,\tau)\rangle_\lambda
   +\langle N_{3(}(t,\tau)\rangle_\lambda= \langle
   N_1(t,\tau)\rangle_\lambda~,\label{X} \\
   Y&=& \langle N_6(t,\tau)\rangle_\lambda
   +\langle N_{4}(t,\tau)\rangle_\lambda=\langle
   N_2(t,\tau)\rangle_\lambda~, \label{Y}
   \end{eqnarray}
where primed quantities ($5'$, $3'$), ($6'$, $4'$) also apply. 
The Bell inequality follows after appropriate
averaging: 
\begin{eqnarray}
\label{Bell_noise}    
&&|F({\bf a,b})-F({\bf a,b^\prime})+
F({\bf a^\prime,b})    +F({\bf a^\prime,b^\prime})|\leq 2~,\\   
&&F({\bf a,b})=
{\langle [N_1({\bf a},t) - N_1(-{\bf a},t)]
[N_2({\bf b},t) - N_2(-{\bf b},t)]\rangle
\over
\langle [N_1({\bf a},t) + N_1(-{\bf a},t)]
[N_2({\bf b},t) + N_2(-{\bf b},t)]\rangle}~,   
\label{Bell_number_F}    
\end{eqnarray}
with ${\bf a}, {\bf b}$ the polarizations of the filters
F$_{1(2)}$ (electrons spin-polarized along ${\bf a}$ (${\bf b}$)
can go through filter F$_{1(2)}$ from lead 1(2) into lead 5(6)).
This is the quantity we want to test, using a quantum mechanical 
theory of electron transport.
Here it will be  written in terms of noise correlators,
as particle number correlators at equal time
can be expressed in general as a function of the 
finite frequency noise cross-correlations. 
Assuming short times (see below), one obtains
$ \langle N_\alpha(\tau)N_\beta(\tau)\rangle\approx \langle
   I_\alpha\rangle \langle I_\beta\rangle \tau^2+\tau
S_{\alpha\beta}(\omega=0)$
where $\langle I_\alpha\rangle$ is the average current in the lead
$\alpha$ and $S_{\alpha\beta}$ denotes the shot noise. 
One then gets:
\begin{eqnarray}   
F({\bf a,b})={S_{56}-S_{54}-S_{36}+S_{34}   
+\Lambda_-\over S_{56}+S_{54}+S_{36}+S_{34}+\Lambda_+}~,   
\label{Bell_noise_F}    
\end{eqnarray}
with
$\Lambda_\pm=\tau(\langle I_5\rangle \pm\langle
I_3\rangle)(\langle I_6\rangle \pm\langle I_4\rangle)$.
Consider now the solid-state analog of the Bell-device as sketched
in Fig. \ref{fig5}b where the particle source is  a
superconductor (S). The test of the Bell inequality
(\ref{Bell_noise}) requires information about the dependence of
the noise on the mutual orientations of the magnetizations
$\pm{\bf a}$ and $\pm{\bf b}$ of the ferromagnetic spin-filters.
 
Consider an example of the solid-state analog of the Bell-device 
[Fig.~1(b)] where the particle source is  a superconductor.
The chemical potential of the superconductor is larger than that
of the leads, which means that electrons are flowing out of the
superconductor. Two 
normal leads 1 and 2 are attached to it in a fork 
geometry
\cite{lesovik_martin_blatter,recher_sukhorukov_loss} and
the filters F$^{\rm e}_{1,2}$  enforce the energy splitting of
the injected pairs. F$^{\rm  d}_{1,2}$-filters play the role of
spin-selective beam-splitters  in the detector. Quasi-particles 
injected into lead 1 and spin-polarized along the magnetization 
$\bf a$ enter the ferromagnet 5 and contribute to the current 
$I_5$, while quasi-particles with the opposite polarization 
contribute to the current $I_{3}$. For a 
biased superconductor with grounded normal leads, 
we find in the tunneling limit the noise   
\begin{equation} 
   \label{S_ideal} 
   S_{\alpha\beta}=
e \sin^2\left(\frac{\theta_{\alpha\beta}}2\right)   
\int_0^{|eV|} d\varepsilon {\cal T}^A(\varepsilon)~, 
\end{equation}  
which integral also represents the current in a given lead
(we have dropped the subscript in ${\cal T}^A(\varepsilon)$ 
assuming the two channels are symmetric). 
Here $\alpha=3,5$, $\beta=4,6$ or vice versa; 
$\theta_{\alpha\beta}$ denotes the angle between the magnetization 
of leads $\alpha$ and $\beta$, e.g., $\cos(\theta_{56})={\bf 
a}\cdot{\bf b}$, and $\cos(\theta_{54})={\bf a}\cdot(-{\bf b})$. 
Below, we need configurations with different settings ${\bf a}$ 
and ${\bf b}$ and we define the angle $\theta_{\bf 
ab}\equiv\theta_{56}$. 
$V$ is the bias of the 
superconductor. 

The $\Lambda$-terms in Eq. (\ref{Bell_noise_F}) 
can be dropped if 
$\langle I_\alpha \rangle \tau\ll 1, $
$\alpha=3,\ldots, 6$, which corresponds to the assumption that
only one Cooper pair is present on average. The resulting BIs 
Eqs. (\ref{Bell_noise})-(\ref{Bell_noise_F}) then neither depend on
$\tau$  nor on the average current but only on the shot-noise,
and  $F=-\cos(\theta_{{\bf ab}})$; the left hand side of Eq.
(\ref{Bell_noise}) has a maximum when $\theta_{\bf ab}=\theta_{\bf 
a^\prime b}=\theta_{\bf a^\prime b^\prime}=\pi/4$ and $\theta_{\bf 
ab^\prime}=3\theta_{\bf ab}$. 
With this choice of angles the BI Eq.(\ref{Bell_noise}) 
is \textit{violated}, thus 
pointing to the nonlocal correlations between electrons in the 
leads 1,2 [see Fig.\ \ref{fig5}(b)]. 
 
If the filters have a width $\Gamma$ the current is of order
$e{\cal T}^A\Gamma/h$ and the 
condition for neglecting the reducible correlators
becomes $\tau\ll \hbar/\Gamma
{\cal T}^A$. 
On the other hand, in order to insure that no electron exchange
between  $1$ and $2$ one requires $\tau \ll \tau_{\rm
tr}/{\cal T}^A$ ($\tau_{\rm tr}$ is the time of flight from
detector $1$ to $2$). The conditions for BI violation require
very small currents, because of the specification that one
entangled pair at a time is in the system. Yet it is necessary
to probe noise cross correlations of these same small currents.
The noise experiments which we propose here are 
closely related to coincidence 
measurements in quantum optics.
\cite{aspect} 

If we allow the filters to have a finite line width, which could
reach the energy splitting of the pair, the violation of BI can
still occur, although violation is not maximal. Moreover, when
the source of electron is a normal source, it is possible to show
that in ``standard device geometries'', where single
electron physics is at play, Bell 
inequalities are not violated: in this situation the term 
$\Lambda_\pm$ dominates over the noise correlation contribution.
Nevertheless, it would be possible in practice to violate Bell
inequalities if the normal source itself, composed of quantum
dots as suggested in Ref. \cite{oliver,loss3}, could generate
entangled electron states as the result of electron-electron
interactions.

Note that there are other inequalities which test entanglement for
particles sources with the number of terminals  two and larger
than (see, e.g., Ref. \cite{werner}); tests of such
inequalities can be implemented in a similar manner as discussed
above. For example, one can use Clauser-Horne (CH) inequalities
\cite{clauser_horne} (instead of Bell Inequalities) to test the
entanglement in the solid-state systems 
shown in the Fig. \ref{fig5}. The
derivation of CH-inequality is similar to that of BI; it is based
on the lemma: if $x,x^\prime,y,y^\prime,X,Y$ are real numbers such
that $x,x^\prime\in [0,X]$ and $y,y^\prime\in [0,Y]$. The 
following inequality then holds:
\begin{equation}
-XY\leq xy-xy^\prime+x^\prime y+x^\prime
y^\prime-Yx^\prime-Xy\leq 0~. 
\end{equation}
The definitions for $x,x',y,y'$ are
the same as in Eqs.(\ref{xx'}) and (\ref{yy'}), but $X=\langle
N_5^{(0)}(t,\tau)\rangle_\lambda$, $Y=\langle
N_6^{(0)}(t,\tau)\rangle_\lambda$, where $\langle
N_6^{(0)}(t,\tau)\rangle_\lambda$ is the number of particles
coming into the arm of the detector $\alpha=5,6$ when it doesn't
include the spin-filter\cite{clauser_horne,Ou}.
Finally we get the CH-inequality in a similar manner as the 
Bell inequality, with the same assumption of small times:
\begin{equation}
\label{S} S_{56}({\bf a,b})-S_{56}({\bf a,b^\prime})+S_{56}({\bf
a^\prime, b})+S_{56}({\bf a^\prime, b^\prime}) -S_{56}({\bf
a^\prime},-)- S_{56}(-,{\bf b})\leq 0~,\nonumber\\
\label{clauser_horne}
\end{equation}
where $S_{56}({\bf a^\prime},-)$ is the shot noise when there is
no spin-filter on the way of the particles coming from the source
into the  the detector arm 6. In the tunneling limit $S_{56}({\bf
a^\prime,b})/S_{56}({\bf a^\prime},-)=\sin^2(\theta_{56}/2)$ and
Eq.~(20) gives $S_{56}$. CH-inequalities are maximally violated
when $\theta_{\bf a,b}=\theta_{\bf a^\prime,b^\prime}=\pi/2$,
$\theta_{{\bf a,b^\prime}}=\pi/4$, and $\theta_{{\bf a^\prime,
b}}=3\pi/4$ [this choice of angles is different than in BI case];
then the left-hand side of Eq. (\ref{S}) is $(\sqrt2 -1)/2$.

CH-inequalities have one working advantage compared to BIs: 
Eq. (\ref{S}) includes only correlations between
terminals 5 and 6,  the number of correlators in Eq. (\ref{S}) is
decreased compared to the BI's. Moreover, in the case of CH
the spin-filters of the detector can include only one ferromagnet 
in each arm rather than
two as in Fig.~5b.


\section{Electron injection in the bulk of a nanotube}

So far, noise was  computed  for non--interacting
systems. If one puts aside the fact that electrons are converted
into holes the calculation of noise in NS systems is similar to
the normal case. Recall now the classic results for the
FQHE.  The long wave length edge excitations along a quantum
Hall bar can be described by a Luttinger liquid \cite{wen}.
Backscattering can be induced by bringing together two
counter--propagating edges using a point contact. In the absence
of  impurities or backscattering, the maximal edge current is
$I_M=\nu e^2/h$, while for weak backscattering, the current
voltage characteristic is highly non linear for Laughlin
fractions, i.e.  $\langle I_B\rangle \sim V^{2\nu-1}$  ($\langle
I_B\rangle$ is the average backscattering current and $V$ is the
voltage bias between the two edges). A
two terminal noise measurement performed on a gated mesoscopic 
device in this regime provides a direct link
to the quasiparticle charge. Quasiparticles are scattered
from one edge to the other one by one, so the usual Schottky
formula $S_B=2e^*\langle I_B\rangle$ which relates the zero
frequency backscattering noise 
to the average current flowing between the two edges
applies \cite{shottky}, with an effective carrier charge
$e^*=\nu e$ contains the electron filling factor
\cite{kane_fisher_noise,chamon_wen,saleur}. This
fractional charge was
measured recently by several groups \cite{saminad,picciotto}.
Statistical interaction between quasiparticle have been addressed
theoretically \cite{safi_devillard_martin}
in an Hanbury-Brown and Twiss
geometry.

Attention is
now turning towards conductors which are essentially free of
defects and which have one--dimensional character. Carbon
nanotubes can have  metallic behavior,
with two propagating modes at the Fermi level, and  constitute
good candidates to study Luttinger liquid behavior. In
particular, their tunnel density of states -- and thus their 
$I(V)$ characteristics -- is known to have a
power law behavior \cite{mceuen,kane_balents_fisher,egger_review}.

\begin{figure}  
\epsfxsize 8 cm  
\centerline{\epsffile{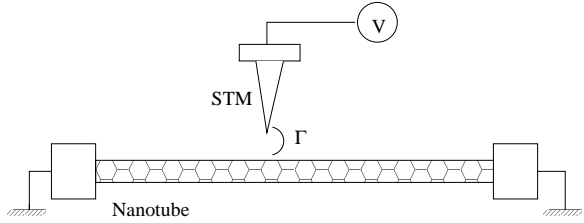}}  
\medskip  
\caption{\label{fig6} Schematic configuration of the
nanotube--STM device: electrons are injected from the tip
at $x=0$: current is measured at both nanotube ends, which
are set to the ground}    
\end{figure} 

Nanotube Luttinger liquids are non-chiral in nature, so
a straightforward transposition of the  results
obtained for chiral edge systems is not obvious. Nevertheless,
non--chiral Luttinger liquids also have underlying chiral 
fields \cite{ines_annales,pham}. Such chiral fields correspond 
to excitations with anomalous (non-integer) charge, which has
eluded  detection so far. 

The transport geometry (Fig.
\ref{fig6}) implies:
tunneling from the tip (normal or ferromagnetic metal)
to the nanotube, and subsequent propagation of collective 
excitations along 
the nanotube. 
In the absence of tunneling, 
the Hamiltonian is thus simply the sum
of the nanotube Hamiltonian, described by a two mode
Luttinger liquid, together with the tip Hamiltonian. Using the
standard conventions \cite{egger_gogolin}, the operator
describing  an electron with spin $\sigma$ moving along the
direction $r$, from mode $\alpha$ is specified  in terms of a
bosonic field:   
\begin{equation} 
\Psi_{r\alpha \sigma}(x,t)={1\over \sqrt{2\pi a}}
e^{i \alpha k_Fx + i r q_Fx + i\sqrt{\frac{\pi}{2}}
\sum_{j\delta}h_{\alpha\sigma
j\delta}(\phi_{j\delta}(x,t)+r\theta_{j\delta}(x,t))}
~,\label{fermion_nanotube}
\end{equation}
with $a$ a short distance cutoff, $k_F$ the Fermi momentum,
$q_F$ the momentum mismatch associated with the two modes, and
the convention $r=\pm$, $\alpha=\pm$ and $\sigma=\pm$ are chosen
for the direction of propagation, for the nanotube branch, and
for the spin orientation. The  non-chiral
Luttinger liquid bosonic fields  $\theta_{j\delta}$ and
$\phi_{j\delta}$, with $j\delta\in\{c+,c-,s+,s-\}$ identifying
the charge/spin  and total/relative fields,
have been introduced. The coefficients are defined as
$h_{\alpha\sigma c+}=1$, $h_{\alpha\sigma
c-}=\alpha$,  $h_{\alpha\sigma s+}=\sigma$ et $h_{\alpha\sigma
s-}=\alpha\sigma$.
The Hamiltonian which describes 
the collective excitations in the nanotube has the standard form:
\begin{eqnarray}
H=\frac{1}{2}\sum_{j\delta}\int_{-\infty}^{\infty}dx\bigg(v_{j\delta}(x)K_{j\delta}(x)
(\partial_x\phi_{j\delta}(x,t))^2
+\frac{v_{j\delta}(x)}{K_{j\delta}(x)}(\partial_x\theta_{j\delta}(x,t))^2\bigg)
~,\end{eqnarray}
with an interaction
parameter $K_{j\delta}(x)$ and velocity $v_{j\delta}(x)$
which allows to address both homogeneous and inhomogeneous
Luttinger liquids.

For the STM tip, one assumes for simplicity that only one
electronic mode  couples to the nanotube, so it can
be described by a semi-infinite  Luttinger liquid
(with interaction parameter $K=1$) for simplicity. 
For the sake of generality, we allow the two spin
components of the tip fields to have different Fermi velocities
$u_F^\sigma$, which allows to treat the case  of a ferromagnetic
metal. The fermion operator at the tip location $x=0$ is then:
\begin{equation}    c_{\sigma}(t)=
\frac{1}{\sqrt{2\pi a}}
e^{i\tilde{\varphi}_{\sigma}(t)}~.
\label{fermion_stm}
\end{equation}
Here, $\tilde{\varphi}_{\sigma}$ is the chiral Luttinger liquid
field, whose Keldysh Green's function (here at
$x=0$) is given in \cite{chamon_freed}.

The tunneling Hamiltonian is a standard hopping term:
\begin{eqnarray}
H_T(t)=\sum_{\varepsilon r \alpha \sigma}
\Gamma^{(\varepsilon)}(t)
[\Psi_{r\alpha\sigma}^\dagger(0,t)c_{\sigma}(t)]^{(\varepsilon)}
~.\label{tunnel_hamiltonian}
\end{eqnarray}
Here the superscript $(\varepsilon)$ leaves either the
operators in bracket unchanged ($\varepsilon=+$), or transforms
them into-- their hermitian conjugate ($\varepsilon=-$). The 
voltage bias between the tip and the nanotube is included using
the Peierls substitution: the hopping amplitude
$\Gamma^{(\varepsilon)}$ acquires a time
dependent  phase $\exp(i\varepsilon \omega_0 t)$, with the bias
voltage  identified as $V=\hbar\omega_0/e$. We
use the convention $\hbar\rightarrow 1$. Similarly, one can
define the tunneling current: \begin{eqnarray}
I_T(t)=ie\sum_{\varepsilon r\alpha\sigma}\varepsilon 
\Gamma_{r\alpha\sigma}^{(\varepsilon)}(t)
[\Psi_{r\alpha\sigma}^\dagger(0,t)c_{\sigma}(t)]^{(\varepsilon)}
~.\end{eqnarray}

For this problem which implies propagation of excitations
along the nanotube  it is also necessary to compute the  (total)
charge current using the bosonized fields Eq.
(\ref{fermion_nanotube}): 
\begin{eqnarray}
I_\rho(x,t)
&=&2ev_F\sqrt{\frac{2}{\pi}}\partial_x\phi_{c+}(x,t)~.
\label{charge_current_operator} \end{eqnarray}
Note that the contribution from terms containing  $2k_F$
oscillations has been dropped. 

The Keldysh technique is employed to compute the 
non-equilibrium currents $\langle I_T \rangle$ and  $\langle
I_\rho(x) \rangle$, and noises $S_T$ and 
$S_\rho(x,x')$ to second order in $\Gamma$. 
The contribution of the nanotube fields 
and of the tip are regrouped into two 
time ordered products. 
Each can be related to an
correlator of several exponentiated bosonic fields.  Such
correlators are readily expressed in terms of the Keldysh Green's
functions $G^{\theta\theta}_{j\delta}$, $G^{\phi\phi}_{j\delta}$,
$G^{\theta\phi}_{j\delta}$, $G^{\phi\theta}_{j\delta}$, associated with the fields  
$\theta_{j\delta}$ and $\phi_{j\delta}$, as
well as the tip Green's function $g_\sigma$.
The following results apply to both homogeneous and inhomogeneous
Luttinger liquids: 
\begin{eqnarray} 
&&\langle I_\rho(x)\rangle=
-\frac{ev_F\Gamma^2}{2\pi^2 a^2}\sum_{\eta\eta_1r_1\sigma_1}
\int_{-\infty}^{+\infty}d\tau \sin(\omega_0\tau)e^{2\pi
g_{\sigma_1 (\eta_1-\eta_1)}(\tau)}\nonumber\\
&&
~~~~~~~~~~~\times
e^{\frac{\pi}{2}\sum_{j\delta}(G^{\phi\phi}_{{j\delta}(\eta_1-\eta_1)}(0,0,\tau)
+G^{\theta\theta}_{{j\delta}(\eta_1-\eta_1)}(0,0,\tau))}
\nonumber\\
&&
~~~~~~~~~~~\times
e^{\frac{\pi r_1}{2}\sum_{j\delta}
(G^{\phi\theta}_{{j\delta}(\eta_1-\eta_1)}(0,0,\tau)
+G^{\theta\phi}_{{j\delta}(\eta_1-\eta_1)}(0,0,\tau))}
\nonumber\\
&&
~~~~~~~~~~~\times\int_{-\infty}^{+\infty}d\tau'
\partial_x\left(G^{\phi\phi}_{c+(\eta\eta_1)}(x,0,\tau')
-G^{\phi\phi}_{c+(\eta-\eta_1)}(x,0,\tau')\right.
\nonumber\\
&&~~~~~~~~~~~~~~~~~~~~~~~
\left.+r_1 G^{\phi\theta}_{c+(\eta\eta_1)}(x,0,\tau')
-r_1G^{\phi\theta}_{c+(\eta-\eta_1)}(x,0,\tau')\right)
~,\nonumber\\
\label{general_nanotube_current}\\
&&S_\rho(x,x',\omega=0)=-\frac{e^2v^2_F\Gamma^2}{(\pi a)^2}\sum_{\eta\eta_1r_1\sigma_1}
\int_{-\infty}^{+\infty}d\tau \cos(\omega_0\tau)
e^{2\pi g_{\sigma_1 (\eta_1-\eta_1)}(\tau)}
\nonumber\\
&&
~~~~~~~~~~~\times
e^{\frac{\pi}{2}\sum_{j\delta}
(G^{\phi\phi}_{j\delta(\eta_1-\eta_1)}(0,0,\tau)+
G^{\theta\theta}_{j\delta(\eta_1-\eta_1)}(0,0,\tau))}
\nonumber\\
&&
~~~~~~~~~~~\times
e^{\frac{\pi r_1}{2}\sum_{j\delta}
(G^{\phi\theta}_{j\delta(\eta_1-\eta_1)}(0,0,\tau)
+G^{\theta\phi}_{j\delta(\eta_1-\eta_1)}(0,0,\tau))}
\nonumber\\
&&
~~~~~~~~~~~\times\int_{-\infty}^{+\infty}d\tau_1
\partial_x\left(G^{\phi\phi}_{c+(\eta\eta_1)}(x,0,\tau_1)
+r_1G^{\phi\theta}_{c+(\eta\eta_1)}(x,0,\tau_1)\right.
\nonumber\\
&&~~~~~~~~~~~~~~~~~~~~~~~
\left.
-G^{\phi\phi}_{c+(\eta-\eta_1)}(x,0,\tau_1)
+r_1G^{\phi\theta}_{c+(\eta-\eta_1)}(x,0,\tau_1)\right)
\nonumber\\
&&
~~~~~~~~~~~\times\int_{-\infty}^{+\infty}d\tau_2\partial_{x'}
\left(
G^{\phi\phi}_{c+(-\eta\eta_1)}(x',0,\tau_2)
+r_1G^{\phi\theta}_{c+(-\eta\eta_1)}(x',0,\tau_2)\right.
\nonumber\\
&&~~~~~~~~~~~~~~~~~~~~~~~
\left.
-G^{\phi\phi}_{c+(-\eta-\eta_1)}(x',0,\tau_2)
+r_1G^{\phi\theta}_{c+(-\eta-\eta_1)}(x',0,\tau_2)\right)
~.\nonumber\\
\label{general_nanotube_noise}
\end{eqnarray}
Here, $\eta,\eta_1,\eta_2=\pm$ are indices which specify on which branch of the Keldysh 
contour the times $\tau,\tau_1,\tau_2$ and $0$ are attached.  

\section{Nanotube noise correlations and effective charges}

An accepted diagnosis to detect effective or anomalous charges
is to compare the noise with the associated current with the
Schottky formula in mind. 
Consider an infinite, homogeneous nanotube characterized by 
interaction parameters $K^N_{j\delta}$. 
The current reads:
\begin{eqnarray}
\langle I_\rho(x)\rangle&=&
\frac{e\Gamma^2}{\pi a}\bigg(\sum_{\sigma}\frac{1}{u_F^{\sigma}}\bigg)
\frac{{\mathrm sgn}(\omega_0)|\omega_0|^{\mu}}{{\bf
\Gamma}(\mu+1)}\bigg(\frac{a}{v_F}\bigg)^{\mu}
{\mathrm sgn}(x)~, 
\label{final_charge_current}\\
\langle I_T(x)\rangle&=& 2|\langle I_\rho(x)\rangle|~,
\end{eqnarray}
where ${\bf \Gamma}$ is the Gamma function.
We thus obtain a non-linear dependence on voltage $|\omega_0|^\mu$ when
interactions are present, with exponent
$\mu=\sum_{j\delta}(K^N_{j\delta}+(K^N_{j\delta})^{-1})/8$.  A
striking result  is that despite the fact that electrons are
tunneling  from the STM tip to the bulk of the nanotube, the
zero frequency current fluctuations are proportional to the
current (for $x'=x>>a$) with an anomalous effective charge:
\begin{eqnarray} S_\rho(x,x,\omega=0)&=&
\frac{1+(K_{c+}^N)^2}{2}e| \langle I_{\rho}(x)\rangle | ~,
\label{auto_noise}\\
S_T&=&e\langle I_T \rangle~.\label{tunnel_noise}
\end{eqnarray} 

More can be learned from a measurement of the noise 
correlations.  
Indeed, our geometry can be considered as a Hanbury-Brown
and Twiss \cite{hbt} correlation device. Such experiments have now been
completed for photons and more recently for electrons in
quantum waveguides. Here the interesting aspect is 
that electronic excitations do not represent the right 
eigenmodes of the nanotube. For $x=-x'>>a$ the noise
cross-correlations read:
\begin{eqnarray}
S_\rho(x,-x,\omega=0)=-\frac{1-(K_{c+}^N)^2}{2}
e|\langle I_\rho(x) \rangle|~.
\label{positive_noise_correlations}
\end{eqnarray}
Note that the prefactors in Eqs. (\ref{auto_noise}) 
and (\ref{positive_noise_correlations}) can readily
be interpreted using the language of Ref. \cite{ines_annales,pham}. 
According to these works, a tunneling event to the bulk of a
nanotube is accompanied by the propagation of two
counter-propagating charges $Q_\pm=(1\pm K_{c+}^N)/2$
in opposite directions. 
Each charge is as likely 
to go right or left. 

The current noise and noise correlations can
be interpreted as  an average over the two types of excitations:
\begin{eqnarray}
S_\rho (x,x) &\sim& {( Q_+^2+Q_-^2 )\over 2} ={1+(K_{c+}^N)^2 \over
4}~,
\label{noise_quasiparticle}\\ 
S_\rho (x,-x) &\sim& -Q_+Q_- =-{1-(K_{c+}^N)^2 \over 4}
~.\label{correlations_quasiparticle}
\end{eqnarray}

The current operator for the nanotube charge is measured in the
same positive $x$ direction at both extremities of the nanotube. 
If one measures the
current away from the  electron source (the STM), 
which corresponds to the standard convention for 
multi-terminal systems,
such correlations become positive. 
Here the noise correlations have the
added particularity  that they occur to second order in 
a perturbative tunneling calculation. 
Strictly positive noise correlations are known to occur 
in superconducting-normal systems with filters, 
with applications toward entanglement 
\cite{lesovik_martin_blatter,recher_loss,bena}.
As noted before, positive correlations do not constitute
a rigorous proof for entanglement. In
the present case, one single electron is injected, but 
it enters a correlated system where electrons are
not ``welcome'' as the eigenstates of the nanotube
consist of a coherent superposition of bosonic modes. 
It therefore has to be 
split into left and right excitations, unless one imposes one
dimensional Fermi liquid leads ($K_{j\delta}^L=1$). Here, we are dealing with 
entanglement between collective excitations 
of the Luttinger liquid. 

\begin{figure}  
\epsfxsize 8 cm  
\centerline{\epsffile{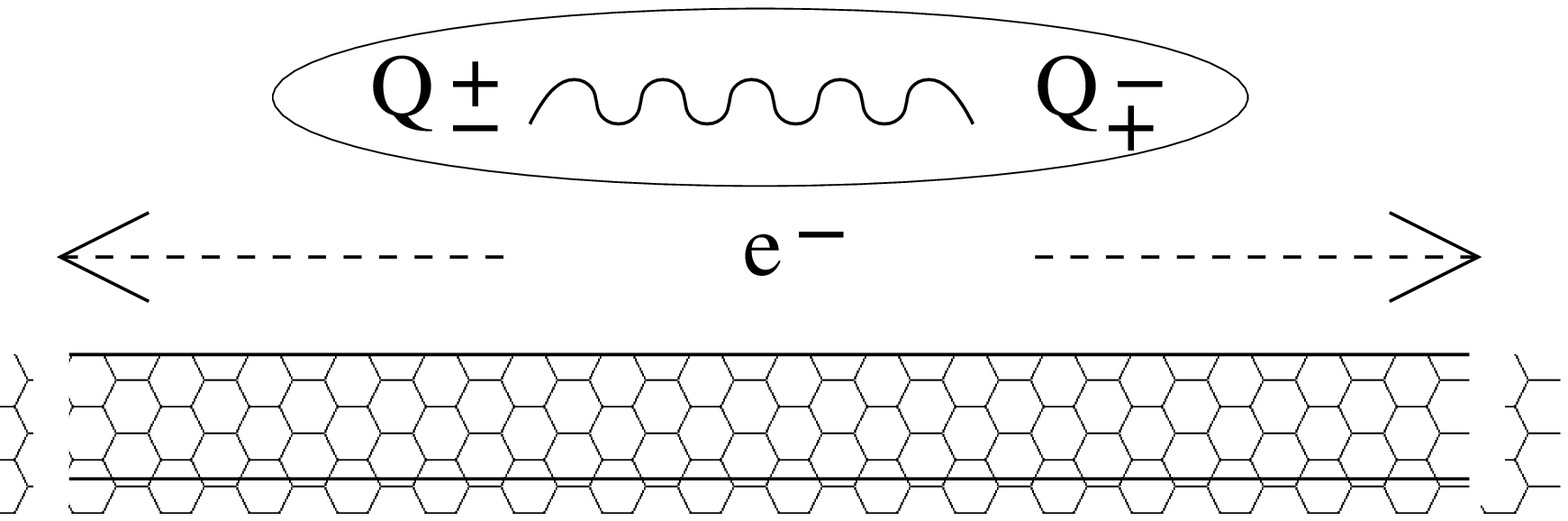}}  
\medskip  
\caption{\label{fig7} An additional electron injected in the bulk of the nanotube 
gives rise to right and left moving chiral excitations which have
entangled charge degrees of freedom.}    
\end{figure}

A drawing where the two types of charges ``flow away''
from the tip while propagating along the nanotube 
is depicted in the lower part of Fig. \ref{fig7}. Both charges $Q_\pm$ are equally
likely to go right or left, and they are emitted as a pair with opposite labels. 
Written in terms of the chiral quasiparticle fields, the addition
of an electron  at $x=0$ with given spin $\sigma$ on a nanotube in the
ground state $|O_{LL}\rangle$ gives: 
\begin{eqnarray}
&&\sum_{r\alpha}\Psi_{r\alpha\sigma}^\dagger(0)|O_{LL}\rangle =
{1\over \sqrt{2\pi a}}\sum_{\alpha}
\prod_{j\delta}
\left\{
[\tilde{\psi}_{j\delta +}^\dagger(0)]^{Q_{j\delta +}}[\tilde{\psi}_{j\delta -}^\dagger(0)]^{Q_{j\delta -}}
\right.
\nonumber\\
&&~~~~~~~~~~~~~~~~~~~~~~~~~~~~~~~~~~~~~~~~~
\left.
+[\tilde{\psi}_{j\delta +}^\dagger(0)]^{Q_{j\delta -}}[\tilde{\psi}_{j\delta -}^\dagger(0)]^{Q_{j\delta +}}
\right\}
|O_{LL}\rangle~,\nonumber\\
\label{entangled_wave_function}
\end{eqnarray}
where for each sector (charge/spin, total/relative mode) the charges
$Q_{j \delta \pm}=(1\pm K_{j \delta}^N)/2$ have been introduced, and 
chiral fractional operators are defined as: 
\begin{equation}
\tilde{\psi}_{j\delta \pm}(x)=
\exp
\left[i 
\sqrt{\frac{\pi }{2 K^N_{j\delta}}}
h_{\alpha\sigma j\delta}\tilde{\varphi}_{j\delta}^\pm(x)
\right]~,
\label{quasiparticle_chiral}
\end{equation}
with $\tilde{\varphi}_{j\delta}^r$ the chiral bosonic fields of
the (nonchiral) Luttinger liquid which time evolution
is simply obtained with the substitution
$\tilde{\varphi}_{j\delta}^r(x)\to\tilde{\varphi}_{j\delta}^r(x-rt)$. 
Consequently, quantum mechanical non-locality is quite explicit
here.
This entanglement is the direct consequence of the correlated
state of the Luttinger liquid: the addition of an electron does 
not yield an eigenstate of the Luttinger liquid Hamiltonian. Electrons must
be decomposed into specific modes, which happen to 
propagate in opposite directions. It 
differs significantly from its  analogs which use
superconductors
\cite{lesovik_martin_blatter,bouchiat,recher_sukhorukov_loss,%
recher_loss,bena}. 

When considering for instance the total charge sector $j\delta=c+$,
the wave function is similar to a triplet spin state
(a symmetric combination of ``up'' and ``down'' states, or
``plus'' and ``minus'' charges) for electrons, with the electrons
being replaced by chiral quasiparticle operators. 
Yet, each chiral field
$\tilde{\varphi}_{j\delta}^r$ can be written as a linear superposition
of boson creation and annihilation operators, and these bosonic 
fields appear in an exponential. This
expresses entanglement between ``many--boson'' states.

We mention briefly the effect of one-dimensional Fermi liquid 
contacts. They can be included as in Ref. \cite{safi_maslov} 
by connecting both nanotube ends to Luttinger liquids
with interaction parameters $K^L_{j\delta}=1$. Standard results 
proper to Fermi liquid systems are
then recovered. To a first approximation, the 
Luttinger liquid parameters of the nanotube 
disappear from transport quantities (current or noise)
when such quantities are evaluated in the leads.
However, higher order corrections in the voltage
still carry a dependence on $K^N_{j\delta}$.
The charge noise follows a Schottky
formula  $S_\rho = e \langle I_\rho \rangle $, and the noise
correlations to order $\Gamma^2$ vanish -- for fermions, 
the leading order being in
$\Gamma^4$. 

\section{Conclusion}

Hanbury--Brown and Twiss geometries provide a
physical test of mesoscopic transport. They can be used to check
the bosonic/fermionic statistics of the carriers, or
alternatively to generate entangled streams of particles.
Information about
statistics is necessarily contained in  quantum measurements
which involve two particles or more: here the zero frequency
noise correlations play the role of  the intensity correlator in
the early quantum optics experiment  of Ref. \cite{hbt}. 

A general form of BI-tests in 
solid-state systems has been proposed, which is formulated in
terms of current-current  cross-correlators (noise correlations), 
the natural
observables in the  stationary transport regime of a solid state
device. For a  superconducting source injecting correlated pairs
into a  normal-metal fork completed with appropriate 
filters \cite{lesovik_martin_blatter,recher_sukhorukov_loss}, the
analysis of such BIs shows that  this device is a source of
entangled electrons with opposite spins
when the fork is  weakly coupled to a
superconductor.  

The possibility of a Bell test for this superconductor-normal
``fork'' devices puts electronic entanglement in condensed 
matter systems on a firm footing. It is now appropriate to imagine 
practical information processing devices which exploit this 
entanglement, in a similar manner as in quantum optics. 
In particular, a proposal for electron 
teleportation consisting of 5 quantum dots
(2 superconducting dots and 3 normal dots) together with a 
superconducting circuit was presented at the workshop.
The spin state of an electron can be transfered to another dot 
without direct matter transfer, and the teleportation sequence is 
selected with the electrostatic interactions between the dots.
Details of this proposal are presented elsewhere 
\cite{sauret_feinberg_martin}.

A diagnosis for detecting the chiral excitations of 
a Luttinger liquid nanotube has been presented, which is based
on the knowledge of low frequency current fluctuation spectrum 
in the nanotube. Typical transport calculations either address
the propagation in a nanotube, or compute tunneling I(V)
characteristics. Here, both is necessary to
obtain the quasiparticle charges. Also, both the noise
autocorrelation and the noise cross-correlations are needed to
identify the charges $Q_{\pm}$. 

Granted, this relies on the
assumption that one dimensional Fermi liquid leads are avoided. 
Multiple scattering at the contacts \cite{imura} may allow to preserve 
the contribution of the noise cross-correlations to second order 
in the tunneling amplitude.
In particular, special
circumstances such as embedded contacts \cite{bouchiat},
transport quantities seem not be renormalized by the contact
parameters. 
This type of geometry could in fact be
tested to analyze the type of contacts which one has
between the nanotube and its connections.  If the ratio $S_\rho
(x,-x,\omega_0)/\langle I_\rho(x)\rangle$ does not depend on the
tunneling distance ($\log \Gamma$), this is a clear indication
that contacts do not  affect this quasiparticle entanglement.

\acknowledgements
The contribution on the Bell inequality tests for electrons
was performed in collaboration 
with G. Blatter and G. Lesovik \cite{chtchelkatchev}; the
contribution on  noise correlations in nanotubes was
achieved together with R. Guyon and P. Devillard
\cite{crepieux}. Discussions and collaborations
with J. Torr\`es, I. Safi and D. Feinberg
on normal-superconducting systems, Luttinger liquids
and teleportation are also gratefully acknowledged. 

 \end{document}